\documentclass[aps,pre,twocolumn,float]{revtex4-1}
\usepackage[normalem]{ulem}
\usepackage{amsmath}
\usepackage{amssymb}
\usepackage{color}
\usepackage{graphicx}
\usepackage{natbib}
\usepackage{CJK}
\usepackage{lipsum}  

\usepackage{xcolor,hyperref}
\hypersetup{
   colorlinks,
   linkcolor={blue!50!black},
   citecolor={blue!50!black},
   urlcolor={blue!80!black}
}

\DeclareMathAlphabet{\mathitbf}{OML}{cmm}{b}{it}

\definecolor{pink}{rgb}{1,0.05,0.7} 
\definecolor{red}{rgb}{1,0,0}  
\definecolor{blue}{rgb}{0,0,1} 
\definecolor{green}{rgb}{0,1,0}
\definecolor{yellow}{rgb}{1,1,0}
\definecolor{orange}{rgb}{1,0.5,0}
\definecolor{white}{rgb}{1,1,1}

\begin{document}

\title{Avalanche properties at the yielding transition: from externally deformed glasses to active systems}

\author{Carlos Villarroel${}^{1}$, Gustavo D\"uring${}^{1,}$${}^{2}$  }
\affiliation{${}^1$Instituto de F\'isica, Pontificia Universidad Cat\'olica de Chile, Casilla 306, Santiago, Chile\\ ${}^2$ANID - Millenium Nucleus of Soft Smart Mechanical Metamaterials, Santiago, Chile}

\date{\today}

\begin{abstract} 
We investigated the yielding phenomenon in the quasistatic limit using numerical simulations of soft particles. Two different deformation scenarios, simple shear (passive) and self-random force (active), and two interaction potentials were used. Our approach reveals that the exponents describing the avalanche distribution are universal within the margin of error, showing consistency between the passive and active systems. This indicates that any differences observed in the flow curves may have resulted from a dynamic effect on the avalanche propagation mechanism. However, we show that plastic avalanches under athermal quasistatic simulation dynamics display a similar scaling relationship between avalanche size and relaxation time, which cannot explain the different flow curves.

\end{abstract}

\maketitle

\section{introduction}
In recent decades, considerable theoretical, experimental, and computational efforts have been made to understand the complex rheology of amorphous materials, such as colloids, grains, foams, and emulsions, which are essential parts of various industrial processes~\cite{apli}. Today, high-density amorphous materials are known to be mechanically stable~\cite{jamming1,jamming2,jamming3,jamming4}; however, they can exhibit an athermal transition between the solid and fluid states when subjected to a sufficiently large shear stress~\cite{yield_exp,yield_exp2,yield_exp3,yield_exp_ch}. These materials exhibit critical stress $\sigma_c$ (sometimes called yield stress). When the applied stress $\sigma\!<\!\sigma_c $, the system moves due to an internal reorganisation, which ends when a configuration capable of bearing the applied stress is found. In this regime, the system behaves as an elastic solid~\cite{buble2}. However, when $\sigma\!>\!\sigma_c$, the system cannot find a stable configuration leading to a flowing state, which is characterised by a singular flow curve relating the strain rate and the shear stress. The flow curve, $\dot{\gamma}\!\sim\!(\sigma-\sigma_c) ^ \beta$, is defined by exponent $\beta$, which is the Herschel-bulkley (HB) exponent~\cite{HB_original}. This dynamic regime is controlled by avalanches composed of several chained irreversible plastic transformations, known as shear transformation zones (STZ), which reorganise a group of particles~\cite{,aval_1,avalan2}. As the flow vanishes, this dynamic becomes increasingly complex, and larger avalanches form, which is reflected in the existence of critical behaviour with a correlation length $\xi\!\sim\! \dot{\gamma}^ {-\nu/\beta} \!\sim\!(\sigma\!-\!\sigma_c)^{-\nu} $ that diverges in $\dot{\gamma}\!\rightarrow\!0$~\cite{avalan2,avalan4}.\\

Yielding-like behaviour is also observed in models of dense active systems subjected to a self-propelled force~\cite{ext_ac_in,active_exp2,active_jam, Ben_Loe,keta2023}. In contrast to systems that exhibit a flow when subjected to sufficient shear stress, the size of the self-propelled force, $f$, must exceed $f_c$. In our previous study~\cite{CD1}, the exponents $\beta$ and $\nu/\beta$ were calculated with good precision for the active and passive scenarios, and they exhibited a difference that did not fall within our range of error. The origin of this difference remains unclear and requires a detailed study of the avalanche statistics and relaxation, which are believed to control the yielding transition. However, obtaining a detailed description of avalanches in a flowing state ($\dot{\gamma}\neq 0$) is a complex task because of the difficulty in detecting and measuring avalanches when the system is not in mechanical equilibrium. \\

\begin{figure}[ht]
\centering
\includegraphics[width = 0.48\textwidth]{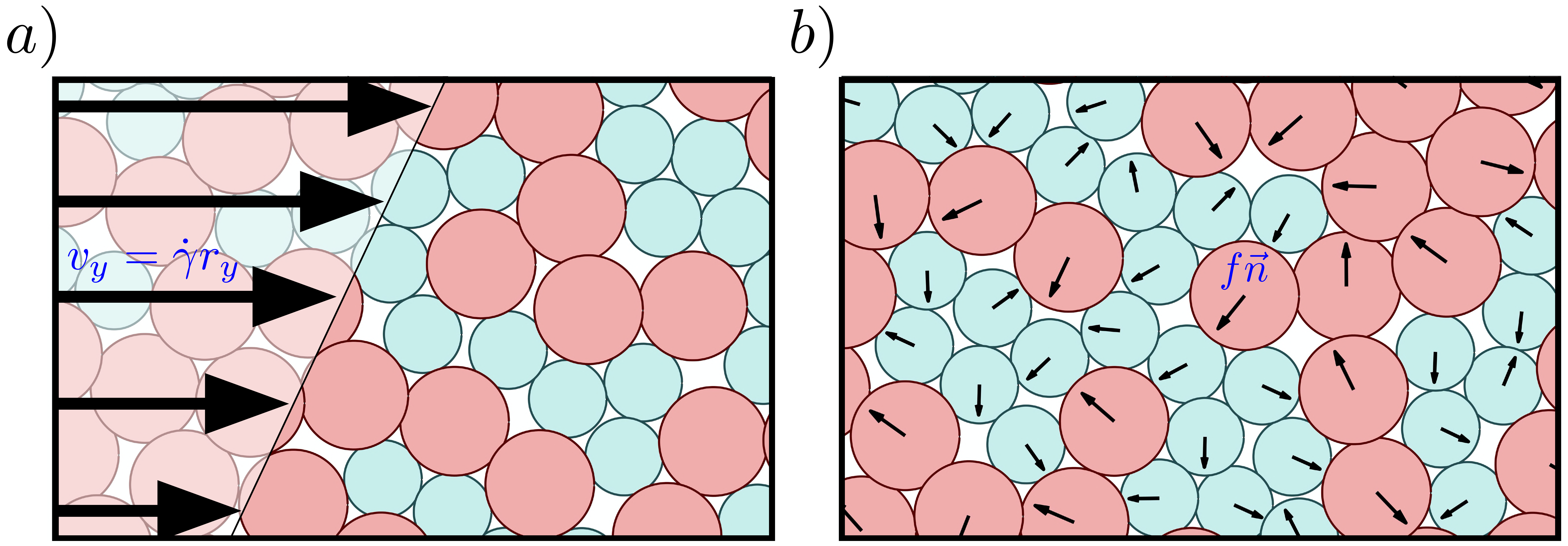}
\caption{\footnotesize a) Representation of the simple shear model, where the system is subjected to a speed profile with equation  $v= \dot\gamma(\vec{r}\cdot\hat{y})\hat{x}$. b) Representation of the self-random force model, where each particle is subjected to a force of size $f$, that is exerted on the direction $\vec{n}^R$. Similar to the simple shear model, when $f>f_c$, the system is not able to find an equilibrium state and flows between non-equilibrium states.}
\label{esq0}
\end{figure}

A common approach to studying the yielding phenomenon and statistics of avalanches in passive systems is to use~\emph{athermal quasistatic simulations} (AQS)~\cite{AQS1,cuasi_buble,LP,AQS2,AQS3,AQS4,AQS5}. This very slow deformation limit is observed when the characteristic time at which the deformation is carried out is sufficient to permit the propagation of avalanches that reorganize the system. This allowed the system to reach a mechanical equilibrium after each deformation step, thereby facilitating the detection of plastic events. In practice, the system is placed under a small and homogeneous strain and then relaxed until mechanical equilibrium is reached. This process is repeated several times until the desired shear strain is achieved~\cite{AQS1}. Although the AQS does not allow a direct study of the fluid regime, this method allows the exploration of the properties and statistics of the avalanche size distribution at the critical point, which controls the dynamics near the critical point~\cite{inertia,AQS1}. In this regard, using mesoscopic elastoplastic models, both Lin et al.~\cite{JieLin_1} and Ferrero et al.~\cite{Ezequiel1} made important advances in matching the exponents that describe the HB rheology with the exponents observed in AQS for simple shear deformation. Nonetheless, AQS for a self-random force is a field that has only recently been studied~\cite{Amiri_2023}. \\


In this study, we performed a large number of simulations involving 2D soft particles in the AQS limit using two distinct models of driven deformation scenarios, \emph{simple shear} (SS) and \emph{self-random force} (SRF), for active particles with infinite persistence in the orientation of self-propulsion, as shown in Fig.~\ref{esq0}. The remainder of this paper is structured as follows: In Sect.~\ref{Num}, detailed information about the simulation protocols utilised throughout this study is provided. In Sect.~\ref{ResAQS}, the results of the avalanche size distributions for both deformation scenarios in the AQS limit are presented. In Sect.~\ref{time_relax}, the propagation times with which the system undergoes reorganisation are examined. Finally, in Sect.~\ref{Disc}, the most significant results are summarised.

\section{Simulation methods and protocols}
\label{Num}
Previous molecular studies can be categorised based on the employed interparticle potential. One group of studies used potentials that diverged when two particles coincided at the centre (e.g., the LJ potential~\cite{inertia,aval_1}, LP potential, and $\sigma/r^{12}$ potential~\cite{LP}). Another group employed potentials in which a finite value is assigned to the same situation (e.g., the Hertzian potential~\cite{jamming5,active_VS_shear} and the harmonic potential~\cite{jamming5,cuasi_buble}). To verify that this selection of potentials does not affect the critical exponents, we employed two potentials: the Hertzian potential and the potential used by Lerner and Procaccia (the LP potential)~\cite{LP}. These potentials were chosen based on the fact that they are differentiable at least twice, thus preventing discontinuity problems in the elastic modulus~\cite{LP}. \\

We used athermal systems of frictionless soft discs for all the simulations in a 2D box of length $L$. To avoid crystallisation, we used a bidisperse mixture of 1:1.4~\cite{cristali_2}. In our simulations, the atomistic length scale was set according to the radius of the small particles ($r_0\!\!=\!\!1$), and the mass of all the particles is equal to unity ($m_0\!\!=\!\!1$). For each potential, the interaction between the particles is described by Eq.~(\ref{Potential_hert}) for the Hertzian potential and Eq.~(\ref{Potential_Edan}) for the LP potential.  

\begin{widetext}
\begin{equation}
U(r_{ij})=\left \{ \begin{matrix} \epsilon \frac{2}{5} \left[1-\frac{r_{ij}}{2d_{ij}}\right]^{5/2}& r_{ij}<2d_{ij}
\\ 0 &  r_{ij}>2d_{ij}  \end{matrix},\right.
\label{Potential_hert}
\end{equation}

\begin{equation}
U(r_{ij})= \left \{ \begin{matrix} \epsilon [ (\frac{d_{ij}}{r_{ij}})^{a_0} - \frac{a_0(a_0 +2 )}{8} (\frac{b_0}{a_0})^{\frac{a_0+4}{a_0+2}} (\frac{r_{ij}}{d_{ij}})^4 + \frac{b_0(a_0+4)}{4} (\frac{r_{ij}}{d_{ij}})^2 - \frac{(a_0+2)(a_0+4)}{8}(\frac{b_0}{a_0})^{\frac{a_0}{a_0+2}} ] & r_{ij}<d_{ij} (\frac{a_0}{b_0})^{\frac{1}{a_0+2}}
\\ 0 &  r_{ij}>d_{ij} (\frac{a_0}{b_0})^{\frac{1}{a_0+2}}  \end{matrix}.\right.
\label{Potential_Edan}
\end{equation}
\end{widetext}

In both cases, $r_{ij}$ is the distance between the centres of particles $i$ and $j$, $d_{ij}$ is the mean of their radii, and $\epsilon$ is the energy scale. In our simulation, we considered $a_0\!=\!10$ and $b_0\!=\!0.2$. The temperature has units of $\epsilon/k_B$, where $k_B$ is Boltzmann's constant, and time is measured in units of $t_0\!=\!\sqrt{m_0r_0^2/\epsilon}$. To create systems with a Hertzian potential, we used infinite quenching~\cite{inf_quench}, and for systems with an LP potential, we started equilibrating at $T\!=\!1.0$ and cooled to $T\!=\!0.05$ at a $10^{-6} \epsilon/(k_B t_0)$ rate; finally, the residual heat was removed using the FIRE algorithm~\cite{FIRE}. Throughout this study, the densities of both potentials were set using the packing fraction $\phi \!=\!0.975$, from the jamming point~\cite{2011APSJammig}. \\

\emph{Quasistatic SS Model}. The AQS for a system under a simple shear deformation can be described using the following procedure~\cite{active_VS_shear,cuasi_buble,AQS1}: Using the Less-Edwards boundary conditions~\cite{lees-edwar}, we imposed an affine shear strain $\Delta \gamma\!=\!10^{-4}$. In each step, we modified the position of each particle $r_i$ according to the following rule:

\begin{equation}
\vec{r}_i \rightarrow \vec{r}_i+ \Delta \gamma (\vec{r}_i \cdot \hat{y}) \hat{x}. 
\label{des_QS}
\end{equation}

After applying the affine deformation, the total potential energy of the system was minimised. To determine the mechanical equilibrium parameters, we defined the residual force factor as  $\lambda_F\!=\!\langle \vec{|F|}  \rangle /\bar{f}$, where $\langle \vec{|F|} \rangle $ is the mean of total force over all particles, $\bar{f}$ is the mean interparticle force, and the mechanical balance is set to $\lambda_F\!<\!10^{-11}$. We primarily used the conjugate gradient (CG) algorithm~\cite{CG-alg} to perform energy minimisation. However, in Appendix~\ref{App_cha2_1}, we tried three energy minimisation methods: FIRE, CG, and Steepest Descent (SD)~\cite{SD}, and we found that this did not affect our results for the occurrence and size of plastic events. The pressure $p$ and shear stress $\sigma$ were quantified following the Irving-Kirkwood calculations~\cite{stress_form}. \\

\emph{Quasistatic SRF Model}.--- A key point in studying the SRF model is identifying quantities and algorithms equivalent to the SS model to be able to make an accurate comparison. Following an overdamped dynamic, the velocity of active particles that are subjected to a persistent self-force with size $f$ can be determined by $\frac{d \vec{r}_i}{d t}  =   D \left[  f \hat{n}^R_i  -  \frac{\partial U}{\partial \vec{r}_i}    \right] $, where $\hat{n}^R_i$ is the direction in which the self-force is applied and $D$ is the overdamp constant~\cite{metodos_SRF,active_exp1}. Despite the simplicity of this approach for conducting simulations, our previous work showed that, to mitigate stagnation issues arising from finite size problems~\cite{CD1}, it is more convenient to reformulate this equation, making the parallel velocity $v^R_\parallel=\frac{1}{N} \sum \vec{v}_i \cdot \vec{n}^R_i$ a control parameter as follows: 

\begin{equation}
\frac{d \vec{r}_i}{d t}  = \frac{L \dot{{\gamma}}^R }{2\sqrt{N}}     \hat{n}^R_i  + D \left[ \bar{f}_\parallel  \hat{n}^R_i  - \frac{\partial U}{\partial \vec{r}_i}    \right].
\label{SRF_di}
\end{equation}

where $\bar{f}_\parallel\!=\!\frac{1}{N} \sum_{j=1}^N \frac{\partial U}{\partial \vec{r}_j}\cdot  \hat{n}^R_j$  represents the mean of the contact force projection along the direction of deformation. In addition, we adopted the definition $\dot{{\gamma}^R}\!=\! \frac{2\sqrt{N}}{L}  v^R_\parallel$, which enabled us to obtain a control parameter with dimensions equivalent to the shear strain rate $\dot{\gamma}$ in the SS model. Consequently, the self-force $f$ is computed as

\begin{equation}
f=\frac{1}{N} \sum_{i=1}^N \left[ \frac{1}{D}\frac{d \vec{r}_i}{d t}  -  \frac{\partial U(r_{ij})}{\partial \vec{r}_i} \right ]\cdot \hat{n}^R_i.
\end{equation}

The final essential ingredient to establish an equivalence between SS and SRF is to define a `random' stress $\sigma^R=\frac{1}{L^2} \frac{dU}{d\gamma^R} =\frac{1}{2 L \sqrt{N}} \sum_{i=1}^N \frac{\partial U}{\partial \vec{r}_i} \hat{n}^R_i$~\cite{active_VS_shear}. By combining this with the overdamped equation, we obtain the following relationship:

\begin{equation}
\sigma^R\!=\! \frac{\sqrt{N}}{2L}f\!-\!\frac{1}{4D}\dot{\gamma}^R.
\end{equation}

In practice, a quasistatic regime is observed when, for $d \gamma$ deformation, the system has enough time to reach a new equilibrium state. Using the above and dynamic Eq.~(\ref{SRF_di}), we constructed a time-independent equation of motion, as described by Eq.~(\ref{AQS-SRF}) that defines how the AQS-SRF should be
\begin{equation}
d \vec{r}_i =  \frac{L }{2\sqrt{N}} d {\gamma}^R     + D \int_{0}^\infty \left[\bar{f}_\parallel   \hat{n}^R_i  - \frac{\partial U}{\partial \vec{r}_i}    \right] dt,
\label{AQS-SRF}
\end{equation}
where  $d {\gamma}^R = \dot{\gamma}^R dt$.\\\
 
Consequently, the AQS algorithm for an SRF deformation can be described as follows: First, at each step of the simulation, an affine deformation displaces the particles in
\begin{equation}
\vec{r}_i \rightarrow \vec{r}_i+ \frac{L}{2\sqrt{N}}     \Delta {\gamma}^R  \hat{n}^R_i. 
\label{des_QS}
\end{equation}
Second, the system is given the time required to reach mechanical equilibrium, which presents a constriction owing to the presence of a self-force $f$. In this sense, minimization is done in search of balance $  \frac{\partial U(r_{ij})}{\partial \vec{r}_i} =   \bar{f}_\parallel \hat{n}^R_i $~\cite{active_VS_shear}. \\

\section{Quasistatic yielding statistics}
\label{ResAQS}
\begin{figure}[h]
\centering
\includegraphics[width = 0.47\textwidth]{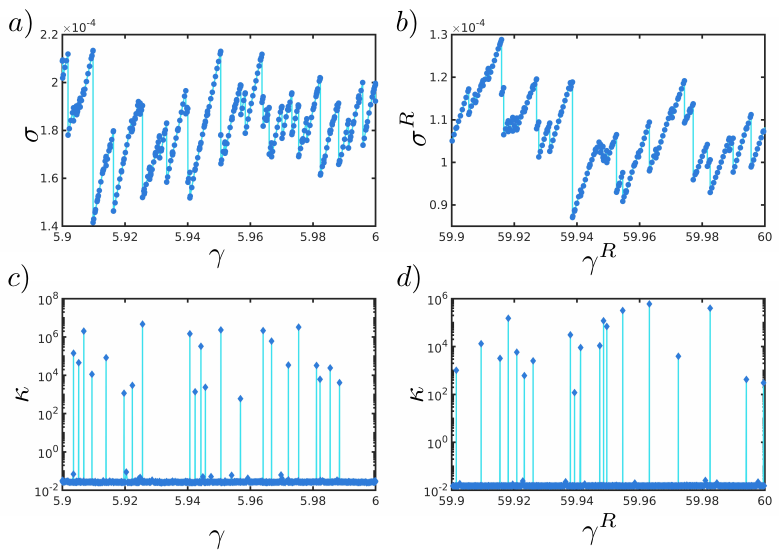}
\caption{\footnotesize Evolution of stress $\sigma$ in a) the SS model and  b) the SRF model as the affine deformation is imposed. c) and d) the evolution of $k$. An abrupt jump is observed when a plastic event occurs. In these data we used potential LP and N=8192.}
\label{esq2}
\end{figure}

As shown in Fig.~\ref{esq2} (a) and (b), we calculate the stress for the SS model over a range of $0\!\!<\!\!\gamma\!<\!\!6$ and an SRF over $0\!\!<\!\!\gamma\!<\!\!60$; using these ranges, we ensured that a stationary state was achieved in the last third of the data, where the data did not depend on the initial configuration. The significant differences in the ranges necessary to obtain these results are consistent with the results obtained in our previous study~\cite{CD1}. \\

To improve the resolution of the plastic events for both models, we used the detection method described by Lerner and Procaccia~\cite{LP}. For each step of size $\Delta \gamma $, this method calculates the difference between the potential energy of the system immediately after making the affine deformation $U_{aff}$ and its energy once it reaches its minimum $U_0$. In Fig.~\ref{esq2}c and d, we show how the total reorganisation factor, $\kappa=\frac{U_{aff}-U_0}{N \Delta\gamma^2}$, fluctuates over the same range as $\gamma$. In intervals in which plastic events are not detected, $\kappa$ assumes values within a well-determined range. However, when an event occurs, this value increases significantly, demonstrating its effectiveness for characterising events with high reorganisation. This allows us to reduce $\Delta \gamma\!=\!10^{-4}$ until $\Delta \gamma\!=\!10^{-6}$ and provides a better resolution in sections close to a plastic event. \\

\begin{figure}[h]
\centering
\includegraphics[width = 0.47\textwidth]{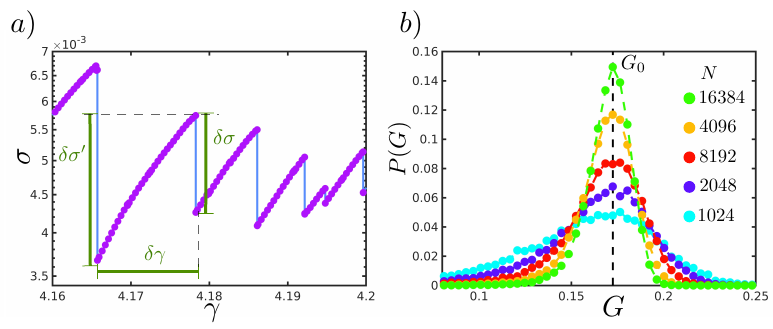}
\caption{\footnotesize a) The system behaves elastically during a section $\delta \gamma$, loading a stress $\delta \sigma'$. This process ends when the system is not able to continue deforming without causing an internal reorganisation (avalanche), which is reflected in a drop in stress $\delta \sigma$. b) The $G$ distribution for different system sizes; Here, the rate at which the system loads stress in each elastic section presents a distribution centred on $G_0$ and diffuses as the size of the system decreases.}
\label{esq3}
\end{figure}

A more detailed analysis of the evolution of $\sigma$ over $\gamma$ reveals that the system exhibits sections with elastic behaviour, where it loads a shear stress $\delta \sigma'$ during a strain section of size $\delta \gamma$ (see Fig.~\ref{esq3}a). This behaviour was also observed when analysing the equivalent quantities for the SRF model. By examining these quantities, we can calculate the shear modulus $G=\frac{\delta \sigma'}{\delta \gamma}$ at which the system loads the stress. In Fig.~\ref{esq3}b, the distribution of $G$ for different system sizes is shown; notably, it follows a distribution centred on $G_0$. 

This elastic section, where the system loads stress, is abruptly terminated by the origin of an avalanche (composed of several plastic events), where the reorganisation of particles occurs. This event is reflected in the  gap in the shear stress of size $\delta \sigma$. \\

\begin{figure}[h]
\centering
\includegraphics[width = 0.48\textwidth]{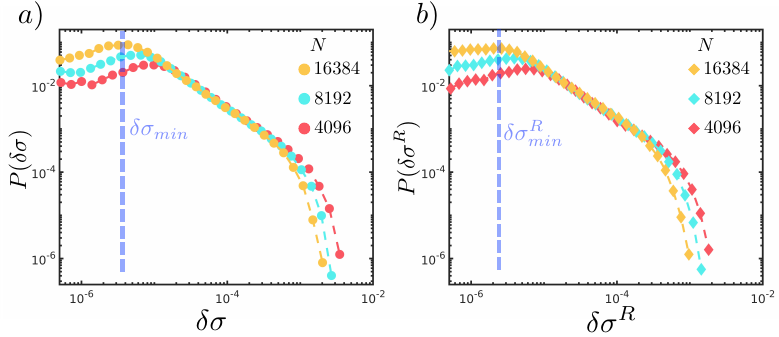}
\caption{\footnotesize a) Distribution of $\delta \sigma$ (SS model) and b) distribution of $\delta \sigma^R$ (SRF model). For $\delta \sigma < \delta \sigma_{min}$ and $\delta \sigma^R < \delta \sigma^R_{min}$, both distributions stop following the power law owing to the presence of a minimum resolution by the simulation protocols.}
\label{esq4}
\end{figure}

Despite the tremendous numerical effort, the $\delta \sigma$ distribution shown in Fig.~\ref{esq4} demonstrates that our data have a minimum resolution $\delta \sigma _\text{min}$, beyond which we cannot capture smaller avalanches. This minimum resolution is the product of $\Delta \gamma$, which is nonzero. In each step, the system loads, on average, a shear stress equivalent to $ G_0 \Delta \gamma$, for which our algorithm that detects the drop in shear stress has problems detecting drops smaller than $G_0 \Delta \gamma$ because the drop in stress can be hidden by the loading process of shear stress. Due to this, and to avoid the diffusion effect of distribution $P(G)$ for small system sizes, we considered only  $\delta \sigma\!> 2 G_0 \Delta \gamma$ to be sufficiently large to avoid minimal resolution problems. \\

The next important result corresponds to the size distribution of an avalanche $S=\delta \sigma L^d$, which is defined as the total stress released by an avalanche in a system of large size $L$. It has been observed that this distribution follows the power law described by $P(S)\sim S^{-\tau}$ and exhibits a cutoff value of $S_c$, which is due to the finite size of the system~\cite{JieLin_1,JieLin_3,AQS2,AQS3,AQS4}. The cutoff corresponds to the size of the system $S_c\sim L^{d_f}$, where $d_f$ is an exponent known as the fractal dimension~\cite{depinning,AQS2}. Together, these two exponents determine the size of the avalanche distribution in the systems. \\

 \begin{figure*}[t]
\centering
\includegraphics[width = 0.98\textwidth]{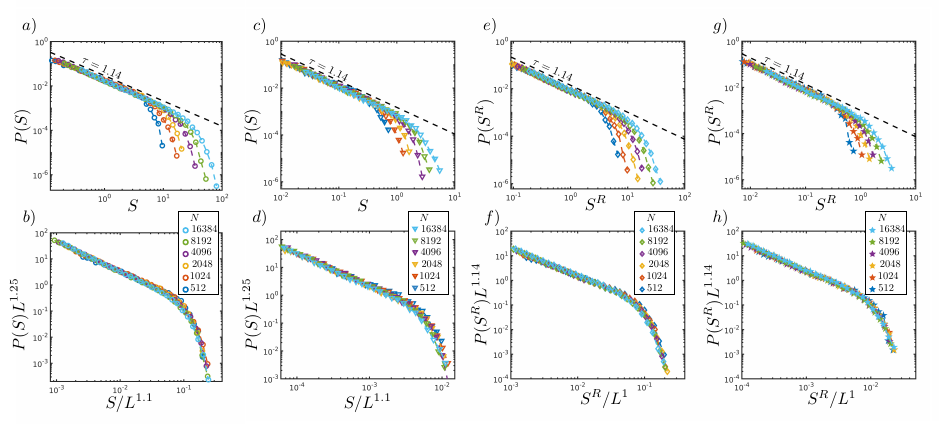}
\caption{\footnotesize a) and c) show the distribution of $S$ for the SS model with LP potential and Hertzian potential for different system sizes, respectively;  e) and g) show the distribution of $S^R$ for the SRF model with LP potential and Hertzian potential for different system sizes, respectively; all results show $\tau\!=\!1.14$. b), d), f), and h) show the same results collapsed by the size of the system $L$; $d_f \!=\!1.1$ for the SS model and $d_f\!=\!1$ for the SRF model.}
\label{esq5}
\end{figure*}

\begin{figure}[h]
\centering
\includegraphics[width = 0.48\textwidth]{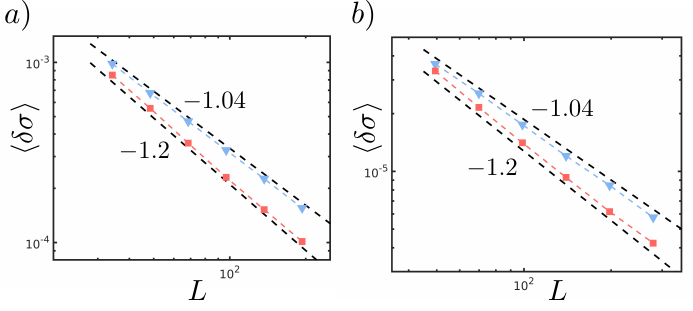}
\caption{\footnotesize The blue line represents $\langle \delta \sigma \rangle$ vs. $L$ based on the established minimal resolution, and the red line $\langle \delta \sigma \rangle$ vs. $L$ represents all plastic events detected for a) the SS deformation and LP potential and b) the Hertzian potential. For both potentials, $\delta = 1.2$ is determined by taking all detected events and $\delta = 1.04$ by adjusting for the presence of a minimum resolution.}
\label{esq6}
\end{figure}

Fig.~\ref{esq5} shows our results for $P(S)$ in both the models and potentials used. We observed that all the distributions have the form $P(S)\sim S^{-\tau} f(S/S_C)$, where $f(x)$ is a rapidly decaying function. A good collapse of the distributions for different system sizes was observed when plotting $P(S)L^{\tau d_f}$ vs. $S/L^{d_f}$~\cite{Ezequiel1}. As can be seen in all configurations used, $\tau\!=\!1.14$ was consistently maintained. However, $d_f$ presents a slight difference between the SS ($d_f\!=\!1.1$) and SRF ($d_f\!=\!1$) models. These values are consistent with those obtained for SS deformation in previous studies~\cite{AQS2,AQS3,AQS4,AQS5}. Using both exponents, we can calculate the scale relation between $ \langle \delta \sigma \rangle $ and the system size $L$ as $ \langle \delta \sigma \rangle \sim  L^{-\delta}$, where $\delta={d - d_f(2-\tau)}$ with $d$ dimension number. The latter result is obtained by integrating $P(S)$ between $0$ and $S_c$. The final scale relation was tested, as shown in Fig.~\ref{esq6} (blue line). Here, $\delta\!=\!1.04$ for the SS model using both potentials, thus reflecting the consistency of the data. However, the $\delta$ exponent that we computed differed from that reported for systems with similar simulation protocols~\cite{LP_res}. This difference is attributable to the fact that our data are truncated for avalanches smaller than $2 G_0 \Delta \gamma$ because, by not differentiating, we recovered exponents similar to those mentioned (red line in Fig.~\ref{esq6}). \\

\section{Avalanche relaxation time in AQS}
\label{time_relax}
The equivalence of results when analysing the avalanche distribution between both deformation scenarios suggests that the discrepancy observed in the exponents describing the fluid region is due to a dynamic component. This dynamic aspect is commonly examined through the exponent $z$, which corresponds to the time required for an avalanche to propagate and its extension length $T\sim l^z$. Additionally, this exponent plays a crucial role in bridging the quasistatic regime with the dynamic regime~\cite{JieLin_1}; however, measuring this exponent has proven to be a challenging task~\cite{JieLin_3,Ezequiel1}. \\

Table~\ref{table1} presents a summary of the exponents calculated for both deformation scenarios. By utilizing these exponents and the relation of the scales studied by Lin et al.~\cite{JieLin_1} on mesoscopic systems ($ \beta =  \nu (d - d_f + z)$), we can indirectly calculate the exponent $z$, resulting in $z=2.9$ for the SS model and $z=8.1$ for the SRF model. Similar to the findings of our previous work~\cite{CD1}, applying this scaling relation indicates that the value of $z$ should be significantly larger than that observed in mesoscopic models~\cite{JieLin_1,Ezequiel1}. \\

\begin{table}[h!]
\centering
\begin{tabular}{||c c c||} 
\hline
Exponent & SS model & SRF model  \\ [0.5ex] 
\hline\hline
$\tau$ & 1.14 & 1.14   \\ 
$d_f$  & 1.1 & 1  \\
$\beta$ & 2.3 & 1.7  \\
$\nu / \beta$ & 0.26  & 0.11 \\ [1ex] 
\hline
\end{tabular}
\caption{\footnotesize Summary of all the exponents calculated throughout this study and our previous study~\cite{CD1} for both deformation scenarios.}
\label{table1}
\end{table}

\begin{figure}[h]
\centering
\includegraphics[width = 0.48\textwidth]{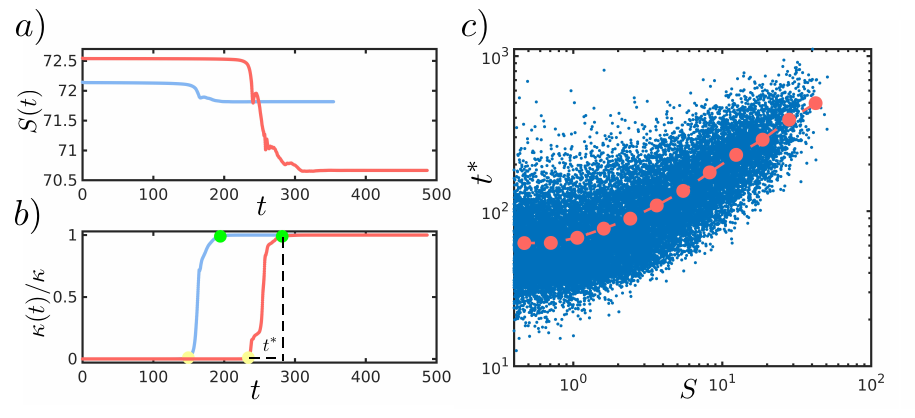}
\caption{\footnotesize  a) S(t) as a function of time $t$ for two different plastic events; these processes end with a total release of stress S. b) $\kappa(t)/k$ as a function of time. Here the time that elapses between the yellow and green point define $t^*$. c) Plot of $t^*$ vs $S$ for one million plastic events in N=8192 systems and LP potential; the red curve shows the average of all events. }
\label{esq7}
\end{figure}

The natural next step is to calculate the exponent $z$ directly. In this context, we propose an algorithm that enables us to make a measurement from AQS. During each simulation step, the system was permitted to relax until it reached a state in which the residual force factor satisfied the equilibrium condition. As mentioned earlier, we measured the total shear stress release $S$ and total reorganisation factor $\kappa$ that occur when a plastic event is generated. Consequently, the time $t^*$ at which this reorganisation process occurs depends on $S$. Figs.~\ref{esq7}a and b illustrate the evolution of $S(t)$ and $\kappa(t)$ over time for two specific events. In the first section (before the yellow dots), the system experiences minimal reorganisation, leading to almost negligible changes in $S(t)$ and $\kappa(t)$. This behaviour is interpreted as a region in which the system is still trying to relax energy following an elastic regime. Similarly, in the last section (after the green dot), the reorganisation is almost negligible, and the system solely aims to adapt to our mechanical stability criteria. In contrast, the central section between the yellow and green dots exhibits a concentration of stress drops and reorganisation. \\

This analysis allowed us to define $t^*$ as the elapsed time in the middle section, which we defined starting at $\kappa(t)/\kappa\!>\!\zeta$ and ending at $\kappa(t)/\kappa\!<\!1-\zeta$, where $\zeta\!=\!0.015$, and verify that variations in this choice do not affect our scale results. In Fig.~\ref{esq7}c), one million events were observed for a system of $N\!=\!8192$ and LP potential.  Here the average of these events indicates that the time needed to reach equilibrium increases in avalanches of larger sizes. \\

\begin{figure}[h]
\centering
\includegraphics[width = 0.45\textwidth]{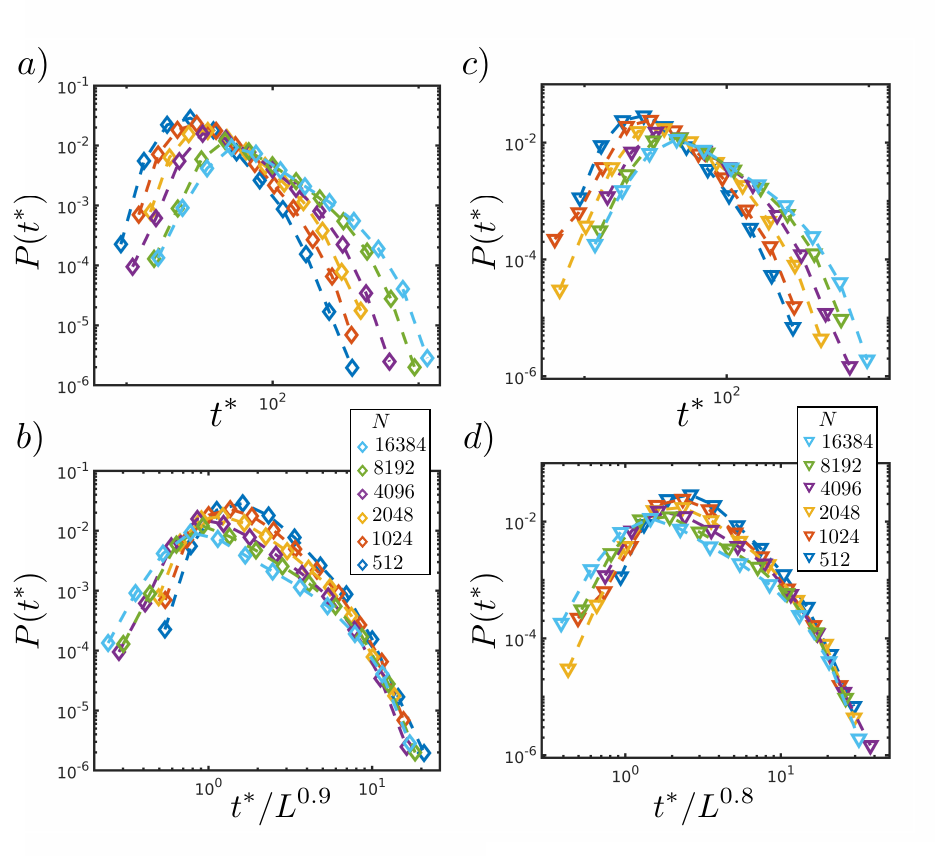}
\caption{\footnotesize a) and b) Distribution $P(t^*)$ for different sizes in the SS and SRF models, respectively. c) and d) the same data collapsed by $L^{0.9}$ and $L^{0.8}$.}
\label{esq8}
\end{figure}

By using this algorithm, Figs.~\ref{esq8}a and b show the distribution $P(t^*)$ for different system sizes with the LP potential method, GC relaxation method, and both deformation scenarios, respectively. The cutoff point of $P(t^*)$ corresponds to the propagation time of an avalanche with a length of the system $L$~\cite{JieLin_1}. Thus, a collapse can be observed with $L^{0.9}$ for the SS model and $L^{0.8}$ for the SRF model. These results provide the exponent $z$, which relates the linear extension of an avalanche vs. the time at which this process occurs using the relation $T\sim L^z$. Here, we obtained $z=0.9$ for passive systems and $z=0.8$ for active systems. This final result represents an important change from the first estimate of the exponent $z$\\

A possible explanation for this radical difference could be the choice of relaxation method used in the simulations. Previous studies on mesoscopic systems found that the value of this exponent is sensitive to the methodology used to propagate the effects of an avalanche~\cite{Ezequiel1}. Similarly, in our soft-particle systems, a point of contention arises regarding the energy minimisation method used in our simulations. The CG algorithms employed in our latest results, or the FIRE algorithms, require significantly less simulation effort than the SD algorithms. Consequently, it is reasonable to expect that this difference will translate into shorter avalanche propagation times for certain relaxation algorithms. Specifically, considering that FIRE incorporates inertia, CG considers the history of descent for faster convergence. A change in the exponent $z$ due to the relaxation method can play an essential role in reconciling the dynamical regime with AQS, especially when considering that almost all studies involving dynamical regimes with particles (including our latest work~\cite{CD1}) use simulation algorithms from Durian's studies~\cite{durian_metodo}, where inertia or any temporary memory effect of the evolution algorithm does not play a role. To address this question, one possible solution would be to perform the same calculation as above for $z$ using the steepest descent (SD) as the relaxation method. However, this requires significant computational effort, as our experience indicates that the simulation times increase between two to three orders of magnitude with SD, making it unfeasible with the current numerical capacity. \\

\section{conclusions}
\label{Disc}
In this study, we observed that the differences in flow curves due to the modification of the deformation scenario type did not appear to be reflected in the avalanche probability distribution when the deformation was executed in a quasistatic regime. This is consistent with previous research that provided similar results~\cite{active_VS_shear,inf_dimen,active_exp1}. \\

Because no difference is observed in the avalanche statistics between the active and passive systems in the quasistatic regime, attention needs to shift towards studying the dynamic properties, specifically the relaxation process of avalanches. For passive systems, a single scaling relation connecting the duration of avalanches with their size has been suggested to link the flow state with $\dot{\gamma}\neq 0$ and avalanche statistics in the AQS regime~\cite{AQS1,inertia,JieLin_1}. This property is characterised by the $z$-exponent introduced in Section~\ref{time_relax}. However, despite its significance, measurements of this exponent in molecular dynamics systems are scarce~\cite{Clemmer2021_1,Clemmer2021_2}. \\

A preliminary measurement of the $z$-exponent under CG dynamics revealed a value much lower than expected, as predicted by the scale relation derived from mesoscopic elastoplastic models~\cite{JieLin_1,Ezequiel1}. However, this difference is probably explained by the relaxation method used in the simulations. \\

\textbf{Acknowledgment.} We thank Edan Lerner for the fruitful discussions and comments on the manuscript. G.D. acknowledges funding from ANID FONDECYT No. 1210656. C.V. acknowledges the support from ANID for Scholarship No. 21181971. \\

\newpage
\bibliography{Biblio_Total}
\appendix
\newpage

\section{$P(S)$ consistency for different relaxation methods}
\label{App_cha2_1}

\begin{figure}[h]
\centering
\includegraphics[width = 0.38\textwidth]{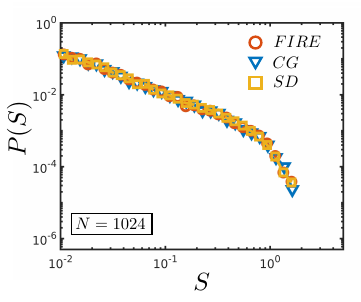}
\caption{\footnotesize  $P(S)$ for a $N=1024$ and the Hertzian potential using three relaxation methods. In red: FIRE, in blue: Conjugate Gradient (GC), and in yellow: Steepest Descent (SD)}
\label{esq15}
\end{figure}

This Appendix verifies that the probability distribution curve $P(S)$ does not vary with the relaxation method used. Fig.~\ref{esq15} shows the results for $P(S)$ for $N=1024$ and the Hertzian potential. The coincidence of the curves can be interpreted as a difference in the relaxation method that did not affect the final equilibrium states.

\end{document}